\documentclass[aps, pra, reprint, superscriptaddress, twocolumn, showkeys, amsmath, amssymb, longbibliography, floatfix]{revtex4-2}
\usepackage[english]{babel}
\usepackage{amsmath, amssymb, bbm, mathrsfs, bm, braket, color, graphicx, comment, amsfonts, dsfont}
\usepackage[colorlinks,citecolor=blue,urlcolor=blue]{hyperref}

\usepackage[utf8]{inputenc}
\usepackage[T1]{fontenc}

\newcommand{\id}{\mathbbm{1}}

\newcommand{\pd}{{\phantom\dag}}

\newcommand{\e}{\mathrm{e}}

\newcommand{\sinc}{\mathop{\mathrm{sinc}}}

\makeatletter
\newcommand*{\transpose}{%
  {\mathpalette\@transpose{}}%
}
\newcommand*{\@transpose}[2]{%
  \raisebox{\depth}{$\m@th#1\intercal$}%
}
\makeatother

\makeatletter
\newcommand*{\blocktranspose}{%
  {\mathpalette\@blocktranspose{}}%
}
\newcommand*{\@blocktranspose}[2]{%
  \raisebox{\depth}{$\m@th#1\intercal$}%
}
\makeatother

\newcommand{\tr}{\mathop{\mathrm{tr}}}

\makeatletter
\DeclareFontFamily{OMX}{MnSymbolE}{}
\DeclareSymbolFont{MnLargeSymbols}{OMX}{MnSymbolE}{m}{n}
\SetSymbolFont{MnLargeSymbols}{bold}{OMX}{MnSymbolE}{b}{n}
\DeclareFontShape{OMX}{MnSymbolE}{m}{n}{
  <-6>  MnSymbolE5
  <6-7>  MnSymbolE6
  <7-8>  MnSymbolE7
  <8-9>  MnSymbolE8
  <9-10> MnSymbolE9
  <10-12> MnSymbolE10
  <12->   MnSymbolE12
}{}
\DeclareFontShape{OMX}{MnSymbolE}{b}{n}{
  <-6>  MnSymbolE-Bold5
  <6-7>  MnSymbolE-Bold6
  <7-8>  MnSymbolE-Bold7
  <8-9>  MnSymbolE-Bold8
  <9-10> MnSymbolE-Bold9
  <10-12> MnSymbolE-Bold10
  <12->   MnSymbolE-Bold12
}{}

\let\llangle\@undefined
\let\rrangle\@undefined
\DeclareMathDelimiter{\llangle}{\mathopen}%
{MnLargeSymbols}{'164}{MnLargeSymbols}{'164}
\DeclareMathDelimiter{\rrangle}{\mathclose}%
{MnLargeSymbols}{'171}{MnLargeSymbols}{'171}
\makeatother

\newcommand{\kket}[1]{\lvert #1 \rrangle}

\newcommand{\bbraket}[1]{\llangle #1 \rrangle}

\usepackage{graphicx}
\usepackage{dcolumn}
\usepackage{bm}
\usepackage{hyperref}
\usepackage[dvipsnames]{xcolor}
\usepackage[mathlines]{lineno}

\begin{document}

\title{Noise resilience of two-dimensional Floquet topological phases}

\author{Balaganchi A. Bhargava}
\affiliation{Institute for Theoretical Solid State Physics, IFW Dresden, Germany and Cluster of excellence ct. qmat, Helmholtzstr. 20, 01069 Dresden, Germany}
\affiliation{Kipu Quantum GmbH, Greifswalderstrasse 212, 10405 Berlin, Germany}

\author{Sanjib Kumar Das}
\affiliation{Institute for Theoretical Solid State Physics, IFW Dresden, Germany and Cluster of excellence ct. qmat, Helmholtzstr. 20, 01069 Dresden, Germany}
\affiliation{Department of Physics, Lehigh University, Bethlehem, PA 18015, USA}
\affiliation{Department of Physics and Astronomy, University of Delaware, Newark, Delaware 19716, USA}

\author{Lukas M. Sieberer}
\affiliation{Institute for Theoretical Physics, University of Innsbruck, 6020 Innsbruck, Austria}

\author{Ion Cosma Fulga}
\affiliation{Institute for Theoretical Solid State Physics, IFW Dresden, Germany and Cluster of excellence ct. qmat, Helmholtzstr. 20, 01069 Dresden, Germany}
\email{i.c.fulga@ifw-dresden.de}

\date{\today}

\begin{abstract}
We study the effect of noise on two-dimensional periodically driven topological phases, focusing on two examples: the anomalous Floquet-Anderson phase and the disordered Floquet-Chern phase. 
Both phases show an unexpected robustness against timing noise. 
The noise-induced decay of initially populated topological edge modes occurs in two stages: 
At short times, thermalization among edge modes leads to exponential decay. 
This is followed by slow algebraic decay $\sim n^{-1/2}$ with the number of Floquet cycles $n$. 
The exponent of $1/2$ is characteristic for one-dimensional diffusion, here occurring along the direction perpendicular to the edge. 
In contrast, localized modes in the bulk exhibit faster decay, $\sim n^{-1}$, corresponding to two-dimensional diffusion. 
We demonstrate these behaviors through full-scale numerical simulations and support our conclusions using analytical results based upon a phenomenological model. 
Our findings indicate that two-dimensional Floquet topological phases are ideal candidates for potential applications of Floquet topology, given the unavoidable presence of both quenched disorder and decoherence in experiments.
\end{abstract}

\maketitle

\section{\label{sec:intro}Introduction}

Non-equilibrium systems obtained via periodic driving have garnered much attention in the previous decades. 
Exploring their topology has been at the forefront of research \cite{Yao2007, Oka2009, Inoue2010, Lindner2011, Lindner2013, Gu2011, Kitagawa2011, Kundu2013, Delplace2013, Katan2013, Lababidi2014, Iadecola2013, Cayssol2013, Goldman2014, Grushin_2014, Kundu2014, Titum2015, Bukov2015} since these systems show phenomena that are not possible in static systems \cite{Kitagawa2010, Rudner2013, Jiang2011, Titum2016, Fulga2016, Asboth2014, Carpentier2015, Nathan2015, Khemani2016, Roy2016, Else2016, Keyserlingk2016, Potter2016, Roy2016, Keyserlingk2016a, Else2016a, Zhang2017, Nathan2017, Kundu2017}.
There have been major experimental advancements in this field \cite{Wang2013, Rechtsman2013, Jotzu2014, Maczewsky2017, Mukherjee2017, Wintersperger2020}, such as realizations of Floquet-Chern phases and of anomalous Floquet topological phases.

While periodic driving opens new avenues towards the generation and manipulation of topology, it brings with it the problem of heating. 
This is well studied in the context of periodically-driven many-body phases, where it has been shown that Floquet systems continuously absorb energy from the driving field, eventually reaching a featureless infinite-temperature steady state \cite{Grushin_2014, D_Alessio_2014}. 
The problem of heating can be (at least partially) overcome if the system is many-body localized \cite{Ponte_2015, Abanin_2016}, or by going to a regime in which the energy absorption rate is exponentially smaller than the driving frequency \cite{Abanin_2015, Abanin_2017}.

Single-particle Floquet systems also heat up, albeit for a different reason: 
In experiments, the driving field is never perfectly periodic. 
The unavoidable deviations from periodicity pose significant limitations to the time scales that are accessible in the lab, especially since the decoherence due to driving noise
is typically exponentially fast \cite{Wintersperger2020}. 
This problem has been acknowledged since the late 1990s, following its experimental observation in quantum kicked rotors \cite{Ammann_1998, Klappauf_1998}. 
Over time, it has evolved into an active research direction, with multiple theoretical \cite{Rieder2018, Sieberer2018, ade2019, Bomantara2020, Ravindranath2020, Timms2021} as well as experimental \cite{Steck2000, dArcy2001, Oskay2003, Sadgrove2004, White2014, Bitter2016, Bitter2017, Sarkar2017, Jrg2017} works studying how noise-induced heating and decoherence occur in single-particle systems, and how it can be mitigated \cite{Selma2021, Nathan2021}, or at least slowed down \cite{Rieder2018, Sieberer2018, Peng2023, Berg2023}.

In this work, we examine the interplay between quenched disorder and noisy time evolution in two-dimensional (2D) Floquet topological phases.  
We focus on one of the paradigmatic toy models, originally introduced by Kitagawa et al.~\cite{Kitagawa2010}, which realizes both a Floquet-Chern phase as well as a so-called anomalous topological phase, in which chiral edge modes form even though the bulk bands have vanishing Chern numbers. 
In each case, we study how a topological edge eigenstate of the Floquet operator decays as a function of time due to noise-induced mixing with other edge and/or bulk modes. 
We find the decay to occur in two stages: 
At short times, the population of the edge state decreases exponentially. 
This first stage of decay lasts until the edge has been thermalized, meaning that all edge modes have been equally populated. 
Remarkably, the edge thermalization time remains finite even in the thermodynamic limit, when the number of edge states goes to infinity. 
Following this initial exponential decay, there is as second stage during which the edge modes decay algebraically with the number of Floquet cycles $n$ as $\sim n^{-1/2}$. 
The second stage lasts until all Floquet eigenstates have equal populations, signalling that the system has heated up to infinite temperature. 
Interestingly, even though these dynamics occur in a 2D system, the exponent of $1/2$ is the hallmark of 1D diffusion. 
This is explained by the fact that after edge thermalization, the population is still localized at the boundary of the system but fully spread along the edge. 
Further spreading can thus occur only along the direction perpendicular to the edge. 
In contrast, localized states in the bulk decay as $\sim n^{-1}$ with an exponent of $1$ as expected in a 2D system. 
The diffusive behavior of edge and bulk states is a consequence of the disorder-induced localization of bulk states, meaning that quenched disorder promotes the resilience of topological modes against decoherence. 
Indeed, the second stage of slow diffusive decay is absent in clean systems with extended bulk states. 
Interestingly, even though spatial disorder cannot fully localize the bulk in the Floquet-Chern phase, our full-scale numerical simulations show that the diffusive behavior is preserved also in this case. 
We provide a heuristic explanation for this result using a simplified phenomenological model that features the fraction of bulk states that are extended as a tuning parameter. 
In this model, diffusive decay persists up to a time scale that scales inversely with that fraction. 
Consequently, this time scale diverges if the number of extended bulk states remains finite in the thermodynamic limit.

The rest of our paper is organized as follows. 
In Sec.~\ref{sec:model} we introduce the toy model of Kitagawa et al.~\cite{Kitagawa2010}, highlighting some of its main features: the existence of different topological phases, as well as the possibility of determining the time-evolution operator analytically \cite{Bhargava2022}, which enables the efficient numerical simulation of large systems. 
Our main result is presented in Sec.~\ref{sec:numerics} where the two different regimes of edge state decay are obtained by means of direct numerical simulations of large systems. 
To get a better handle on the mechanism governing the system's behavior, we turn in Sec.~\ref{sec:superop} to a superoperator formalism, which enables us to average over noise realizations analytically, while obtaining results consistent with the full-scale numerical simulations.
Using this formalism, we show that the change from exponential to algebraic decay is associated with a thermalization of the edge states, separate from the bulk modes. 
In Sec.~\ref{sec:pheno-model}, we introduce a phenomenological model that explains the observed edge thermalization. 
Using a semiclassical rate equation, we demonstrate that the edge thermalization time remains finite even in the thermodynamic limit, when the number of edge modes goes to infinity. 
Furthermore, the phenomenological model offers a heuristic explanation for the persistence of diffusive decay in the presence of extended bulk states in the Floquet-Chern phase. 
We conclude in Sec.~\ref{sec:conclusions}. 
All of our data and the code used to generate it are available on Zenodo~\cite{Zenodo_code}.

\section{Model}
\label{sec:model}

\begin{figure*}[tb]
\centering
\includegraphics[width=0.85\textwidth]{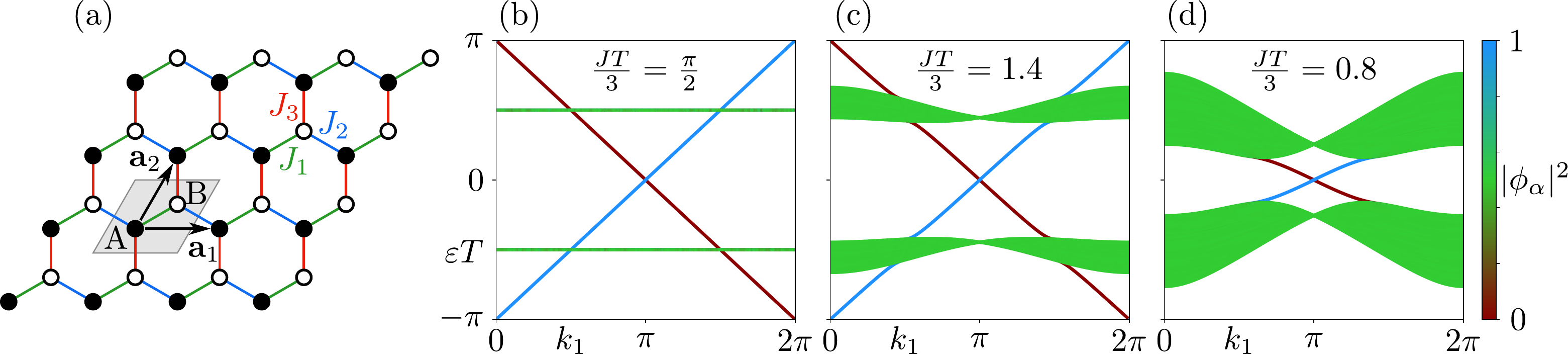}
\caption{
Panel (a) shows a sketch of the tight-binding model. 
The unit cell (gray box) contains two sites belonging to the A and B sublattices, marked as filled and open circles, respectively. 
The Bravais vectors are ${\bf a}_1$ and ${\bf a}_2$. 
The three different nearest-neighbor hoppings, $J_1$, $J_2$, and $J_3$, are shown using green, blue, and red colors, respectively. 
The other panels show the bandstructure (quasi-energy $\varepsilon T$ versus dimensionless momentum $k_1$) of the system in a ribbon geometry, infinite along the ${\bf a}_1$ direction and consisting of 20 unit cells along the ${\bf a}_2$ direction. 
The color scale denotes the probability density summed over the bottom-most 10 unit cells of the ribbon, such that bulk states are shown in green, whereas states localized on the top/bottom edges are shown in red/blue, respectively. 
We use values of $JT/3=\pi/2$ in panel (b), $1.4$ in panel (c), and $0.8$ in panel (d).
}
\label{fig:model}
\end{figure*}

We consider the model first used by Kitagawa et al.~\cite{Kitagawa2010} to study the topological properties of Floquet systems. 
It consists of spinless fermions hopping on a honeycomb lattice, where the hopping integrals are varied periodically in time.
There are two sites per unit cell, belonging to sublattices A and B, and the Bravais vectors ${\bf a}_1$ and ${\bf a}_2$ are as shown in Fig.~\ref{fig:model}.
The momentum-space Hamiltonian reads:
\begin{equation}\label{eq:Hk}
 H = \sum_{\bf k} 
 \begin{pmatrix}
  c_{A,{\bf k}}^\dag & c_{B,{\bf k}}^\dag
 \end{pmatrix}
 \begin{pmatrix}
  h_{AA} & h_{AB} \\
  h_{BA} & h_{BB}
 \end{pmatrix}
 \begin{pmatrix}
  c_{A,{\bf k}}^\pd \\
  c_{B,{\bf k}}^\pd
 \end{pmatrix}
\end{equation}
with $h_{AA}=h_{BB}=0$ and
\begin{equation}
 h_{AB}^\pd = h_{BA}^* = J_1 + J_2 e^{-i k_1} + J_3 e^{-i k_2}.
\end{equation}
Here, $c_{A/B,{\bf k}}^\dag$ are creation operators on the two sublattices, whereas ${\bf k} = (k_1, k_2)$ are the two dimensionless momenta pointing along the two Bravais vectors ${\bf a}_{1,2}$.

We make the Hamiltonian time dependent by varying the nearest-neighbor hopping amplitudes $J_{1,2,3}$ (shown in Fig.~\ref{fig:model}) in a cyclic manner. 
The driving protocol consists of three steps, during each of which only one of the three hopping amplitudes is nonzero. 
Thus, during the $n^\text{th}$ driving cycle we have:
\begin{enumerate}\label{eq:driving_protocol}
\item $J_1 = J$, $J_{2}=J_3=0$ 

for $\left( n - 1 \right) T<t\leq \left( n - 1 \right) T+ T/3$,\\
\item $J_2 = J$, $J_{1}=J_3=0$ 

for $\left( n - 1 \right)T+T/3<t\leq \left( n - 1 \right)T+ 2T/3$,\\
\item $J_3 = J$, $J_{1}=J_2=0$ 

for $\left( n - 1 \right)T+2T/3<t \leq nT$,
\end{enumerate}
where $t$ is time, $T$ is the driving period, and $J$ denotes the strength of the nonzero hopping. 
Since the Hamiltonian is piecewise constant as a function of time, the time-evolution operator over one full driving cycle, also referred to as the Floquet operator, reads:
\begin{equation}\label{eq:Floquet}
F = e^{-iH_{3}\frac{T}{3}}e^{-iH_{2}\frac{T}{3}}e^{-iH_{1}\frac{T}{3}},
\end{equation}
where we work in units such that $\hbar=1$, and where $H_{1,2,3}$ are the Hamiltonians in each of the three steps of the driving cycle. 
Since at each point in time the system is composed of dimers that are decoupled from each
other, the Floquet operator can be computed analytically.

This model realizes different Floquet topological phases as a function of $JT$, the dimensionless product between the hopping strength and the period. 
To see this, we consider a ribbon geometry, in which the system has translation symmetry along the ${\bf a}_1$ direction, but is finite along ${\bf a}_2$. 
We diagonalize the Floquet operator as
\begin{equation}
  \label{eq:Floquet-states}
F \ket{\phi_\alpha} = e^{-i\varepsilon_\alpha T} \ket{\phi_\alpha},
\end{equation}
where $\{ \ket{\phi_\alpha} \}$ is the orthonormal basis of Floquet eigenstates, and plot its $2\pi/T$ periodic eigenphases $\varepsilon_\alpha$, also referred to as quasi-energies as a function of momentum in Fig.~\ref{fig:model}(b-d).
For $JT/3=\pi/2$ and $1.4$ (panels b and c), the model hosts a so-called anomalous Floquet topological phase, one in which chiral edge modes (shown in red and blue) coexist with topologically trivial bulk bands (green).  
This is possible because of the periodicity of the quasi-energy Brillouin zone, which enables edge modes to exist both in the $\varepsilon T =0$ and $\varepsilon T = \pi$ quasi-energy gaps. 
Note that for $JT/3=\pi/2$, which we refer to as the resonant driving point, the bulk bands are dispersionless; 
they acquire a finite bandwidth away from this point.

For $J T/3 = 0.8$, on the other hand, there are no chiral edge states in the $\varepsilon T = \pi$ gap [see Fig.~\ref{fig:model}(d)], and the bandstructure resembles that of a conventional, static Chern insulator, e.g. the Haldane model \cite{Haldane_1988}.
This is the so-called Chern phase, in which the presence of edge modes is tied to the Chern numbers of the bands, the latter taking the values $\pm 1$.

\section{Timing noise}
\label{sec:numerics}

We model deviations from periodic driving by randomizing the duration of each step in the driving cycle. 
Thus, the time-evolution operator corresponding to the $n^\text{th}$ cycle becomes
\begin{equation}\label{eq:noisy_cycle}
 U_n = e^{-iH_{3}\frac{T_{n, 3}}{3}}e^{-iH_{2}\frac{T_{n, 2}}{3}}e^{-iH_{1}\frac{T_{n, 1}}{3}},
\end{equation}
where $T_{n, m} = T + \tau_{n, m}$, and $\tau_{n, m}$ are randomly drawn from the uniform distribution $[-V_t, V_t]$, such that $V_t$ is the noise strength.
From here on, we will consider weak noise strength, setting $J V_t / 3 = 0.05$.

Since time periodicity and thus quasi-energy conservation are now broken, we no longer work with a Floquet operator, but with the full time-evolution operator corresponding to $n$ driving cycles,
\begin{equation}\label{eq:noisy_driving_cycles}
  U_{n, 1} = U_n U_{n-1} \dotsb U_2 U_1.
\end{equation}
Our strategy is to consider an initial state, labeled $\ket{\psi_0}$, which is chosen to be an edge eigenstate of the noiseless Floquet operator. 
According to Eq.~\eqref{eq:Floquet-states}, such a state would be left invariant (up to a phase factor) under perfectly periodic time evolution. 
We compute the population of this state as the system undergoes multiple cycles of noisy driving [c.f. Eq.~\eqref{eq:noisy_driving_cycles}]:
\begin{equation}
  \label{eq:pop-one-seed}
 \mathcal{P}_n = \lvert \braket{\psi_0|\psi_n} \rvert^2 = \lvert \braket{\psi_0|U_{n,
     1}|\psi_0} \rvert^2.
\end{equation}

\subsection{Clean case}

Note that introducing noise as per Eqs.~\eqref{eq:noisy_cycle} and \eqref{eq:noisy_driving_cycles} does not break the translation symmetry of the ribbon discussed before, meaning that the momentum $k_1$ remains conserved even when quasi-energy conservation is broken. 
Thus, studying noise can be done by considering a set of 1D problems, one for each initial state $\ket{\psi_0 (k_1)}$.

\begin{figure}[tb]
\includegraphics[width=0.8\columnwidth]{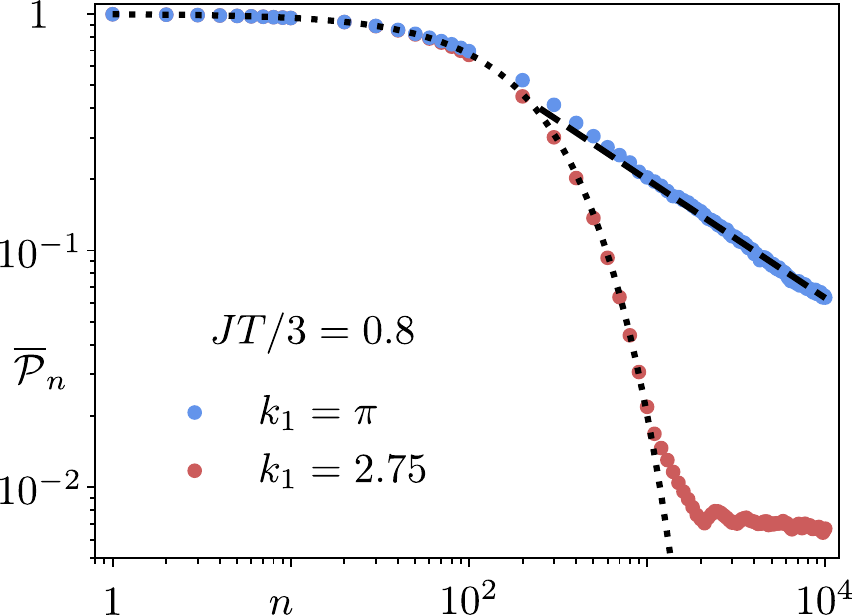}
\caption{
The noise-averaged edge state population is plotted as a function of number of cycles for $JT/3=0.8$ and $JV_t/3=0.05$, in the case of an initial state localized at the bottom boundary of a ribbon consisting of 100 unit cells. 
The blue dots correspond to the state at $k_1=\pi$, whereas the red dots are for $k_1=2.75$.
The population is averaged over 5000 noise realizations, with error bars smaller than symbol sizes. 
The dashed line shows an algebraic decay $\sim n^{-1/2}$. 
The dotted line shows an exponential decay. 
When $k_1=2.75$, the flat region at large $n$ is a finite-size effect: 
The initial state has already spread out over all sites in the system.
}
\label{fig:clean_ribbon}
\end{figure}

The results we obtain in the ribbon geometry are identical to those previously reported in Refs.~\cite{Rieder2018, Sieberer2018}: 
The initial edge mode (taken to be on the bottom edge of the ribbon) decays exponentially whenever the bulk bands have a finite bandwidth (at fixed $k_1$), whereas it decays algebraically, $\mathcal{P}_n \sim n^{-1/2}$ with the exponent of $1/2$ being the hallmark of 1D diffusion, when bulk bands are flat. 
Therefore, at $JT/3=\pi/2$ [see Fig.~\ref{fig:model}(b)] all initial edge states $\ket{\psi_0 (k_1)}$ exhibit diffusive behavior. 
Interestingly, away from this resonant driving point, exponential decay occurs for all initial edge states except for those at $k_1=\pi$. 
At this momentum all bulk states are concentrated at two quasi-energies, see Fig.~\ref{fig:model}(c, d). 
Thus, the 1D effective system obtained by fixing $k_1=\pi$ is characterized by dispersionless 1D bulk states, which have been shown to result in algebraic decay. 
These results are summarized in Fig.~\ref{fig:clean_ribbon}, which shows the noise-averaged population $\overline{\mathcal{P}}_n$ as a function of the number of noisy driving cycles, $n$, for initial states at different $k_1$.

References~\cite{Rieder2018, Sieberer2018} provide the following intuitive explanation for the behavior of the edge mode population in 1D: 
In the limit of weak noise, one can think of time evolution as proceeding for multiple cycles according to the perfectly-periodic driving protocol, with occasional instances of imperfect driving cycles, or `errors.' 
Each of these events leads to a mixing of states at different quasi-energies, thus allowing part of the edge mode population to be transferred to bulk states. 
In the case when the 1D bulk states are dispersionless, this results in a 1D random walk between the initial edge mode and bulk modes that are localized close to the boundary, leading to diffusive decay, $\sim n^{-1/2}$. 
In contrast, when bulk states are extended, any newly populated state travels away from the edge with a finite group velocity, resulting in an exponential decay of the initial edge mode population.

\subsection{Disordered case}
\label{sec:disordered-case}

We now move to genuinely 2D systems and examine the interplay between quenched disorder and noise. 
To this end, we consider the model in a cylinder geometry (periodic boundary conditions in the ${\bf a}_1$ direction) with a circumference of $L$ unit cells and a height of $W$ unit cells. We introduce onsite disorder in the system, with a real-space disorder Hamiltonian
\begin{equation}
 H_{\text{dis}} = \sum_{s, i, j} \mu^\pd_{s,i,j} c^\dag_{s,i,j} c^\pd_{s,i,j},
\end{equation}
that is added to all three Hamiltonians $H_{1,2,3}$ in Eqs.~\eqref{eq:Floquet} and \eqref{eq:noisy_cycle}. 
In the disorder Hamiltonian, $s=A,B$ denotes the sublattice, $i, j$ are the indices of a particular unit cell, and the disordered onsite potential $\mu_{s,i,j}$ is drawn independently for each lattice site from the uniform distribution $[-V_o, V_o]$, with $V_o$ the onsite disorder strength.

\begin{figure}[tb]
\includegraphics[width=0.8\columnwidth]{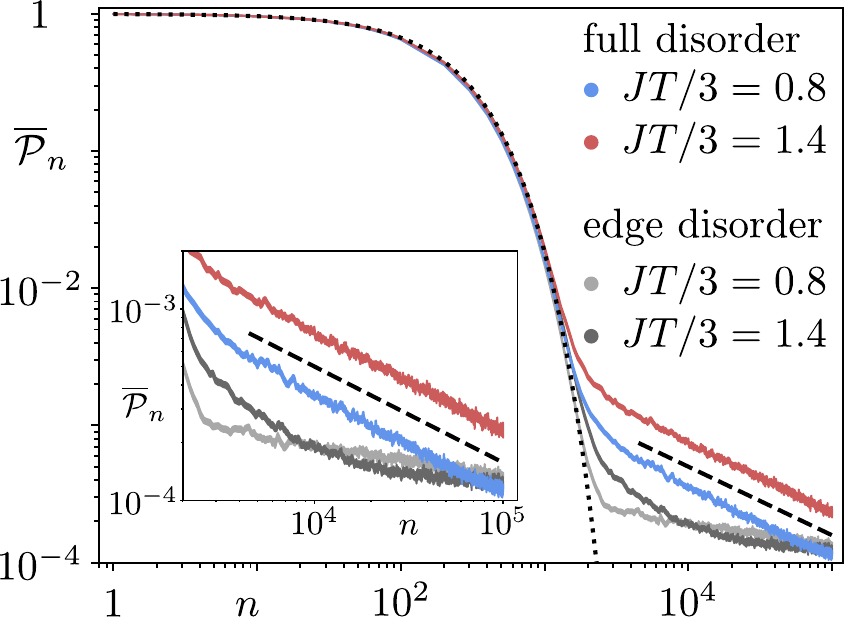}
\caption{
The edge state population decay for the anomalous ($JT/3=1.4$, red) and Chern ($JT/3=0.8$, blue) phases. 
We use $JV_t/3=0.05$ and $V_oT/3=0.3$. 
The disordered cylinder has a circumference of $L=24$ unit cells and a height of $W=200$ unit cells.
The population is averaged over 1000 noise realizations, with error bars indicated by line thickness.
The red and blue solid lines show results obtained when disorder is added to all sites, whereas the gray lines are obtained when adding disorder only to sites on the edge of the cylinder.
The dotted line shows an exponential decay, whereas the dashed one shows an algebraic decay, $\sim n^{-1/2}$. 
The inset is a closeup of the diffusive region.
}
\label{fig:disordered_cylinder}
\end{figure}

We set $V_oT/3=0.3$ and choose as initial state the eigenstate of the Floquet operator with quasi-energy closest to $0$, which we verify to be an edge mode localized on either the top or the bottom of the disordered cylinder. 
The noise-averaged population as a function of the number of driving cycles is shown in Fig.~\ref{fig:disordered_cylinder}, both for $JT/3=1.4$ (anomalous phase) and for $JT/3=0.8$ (Chern phase).

The population of the edge mode (red and blue lines in Fig.~\ref{fig:disordered_cylinder}) shows a behavior that is a mix between those observed in the 1D case (cf. Fig.~\ref{fig:clean_ribbon}).
The initial decay is exponential until $n \simeq 3000$ driving cycles, after which it becomes algebraic.
Notice that this occurs both in the case of the anomalous Floquet phase as well as in the disordered Chern phase, and that it is associated with the presence of bulk disorder.
When disorder is added only to the edge sites, such that all bulk states remain extended, no diffusive decay is observed, as indicated by the light and dark gray lines in Fig.~\ref{fig:disordered_cylinder}.
Instead, the population decays exponentially up until the initial state has spread over the entire cylinder, leading to the flat regions at large $n$, which are due to finite-size effects.

Interestingly, we note that the presence of bulk disorder affects both the anomalous phase and the Chern phase in the same way, even though their bulk localization properties are different. 
For the former, bulk bands are trivial, and will thus be fully localized by disorder \cite{Titum2016}. 
In the case of the Chern phase however, not all bulk states can be localized \cite{Halperin1982, Thouless1984, Thonhauser2006}: 
There must remain at least one delocalized mode in each bulk band, which ``carries the Chern number.''

\section{Superoperator formalism}
\label{sec:superop}

The average over noise, which we have performed numerically to obtain the results presented in the previous Sec.~\ref{sec:numerics}, can also be carried out analytically using the superoperator formalism~\cite{Rieder2018, Sieberer2018, Timms2021}. 
This enables the efficient calculation of the populations of not only the initial edge state but of all eigenstates of the noiseless Floquet operator, and thus leads to a better understanding of the dynamics. 
Moreover, the Floquet superoperator forms the basis for the phenomenological model introduced in Sec.~\ref{sec:pheno-model} below, which offers a simplified heuristic framework to obtain analytical explanations for our results.

To derive the Floquet superoperator that describes the noise-averaged dynamics, we first consider the projector onto the pure state of the system after $n$ noisy driving cycles, $\ket{\psi_n}\bra{\psi_n}$. 
During the next driving cycle, the projector evolves as
\begin{multline}
  \label{eq:projector-one-cycle-evolution}
  \ket{\psi_{n + 1}} \bra{\psi_{n + 1}} \\ = e^{- i \mathcal{H}_{3} \frac{T_{3,
        n + 1}}{3}} e^{- i \mathcal{H}_{2} \frac{T_{2, n + 1}}{3}} e^{- i
    \mathcal{H}_{1} \frac{T_{1, n + 1}}{3}} \ket{\psi_n}\bra{\psi_n},
\end{multline}
where we have introduced the Hamiltonian superoperators $\mathcal{H}_{1, 2, 3}$, which are defined in terms of the ordinary Hamiltonians $H_{1, 2, 3}$ that generate the evolution during the three steps of the cycle as
\begin{equation}
  \label{eq:H-m-superoperator}
  \mathcal{H}_m \rho = [H_m, \rho], \qquad m = 1, 2, 3.
\end{equation}
Now we take the average over noise, which leads to the density matrix $\rho_n = \overline{\ket{\psi_n} \bra{\psi_n}}$. 
The evolution equation for the density matrix follows directly from Eq.~\eqref{eq:projector-one-cycle-evolution} and by using the fact that the noise in each driving cycle is statistically independent,
\begin{equation}
  \label{eq:Floquet-superoperator-evolution}
  \rho_{n + 1} = \mathcal{F} \rho_n,
\end{equation}
where the Floquet superoperator is given by
\begin{equation}
  \label{eq:Floquet-superoperator}
  \begin{split}
    \mathcal{F} & = \overline{e^{- i \mathcal{H}_{3} \frac{T_{3, n + 1}}{3}}
      e^{- i \mathcal{H}_{2} \frac{T_{2, n + 1}}{3}} e^{- i \mathcal{H}_{1}
        \frac{T_{1, n + 1}}{3}}} \\ & = e^{- i \mathcal{H}_{3} \frac{T}{3}}
    \mathcal{E}_{3} e^{- i \mathcal{H}_{2} \frac{T}{3}} \mathcal{E}_{2} e^{- i
      \mathcal{H}_{1} \frac{T}{3}} \mathcal{E}_{1}.
  \end{split}
\end{equation}
In this expression, the factors $e^{- i \mathcal{H}_m \frac{T}{3}}$ describe the noiseless evolution. 
The superoperators $\mathcal{E}_m$ result from averaging over the random time shifts $\tau \in [-V_t, V_t]$,
\begin{equation}
  \label{eq:noise-superop}
  \mathcal{E}_m = \overline{e^{-i \mathcal{H}_m \frac{\tau}{3}}} =
  \frac{1}{2V_t} \int_{- V_t}^{V_t}d\tau \, e^{- i\mathcal{H}_m
    \frac{\tau}{3}} = \sinc \! \left( \mathcal{H}_m
    \frac{V_t}{3} \right),
\end{equation}
where $\sinc(x) = \sin(x)/x$. 
As detailed in Appendix~\ref{appx:numerical-superoperator}, we implement the Floquet superoperator $\mathcal{F}$ numerically by rewriting Eq.~\eqref{eq:Floquet-superoperator-evolution} in vectorized form: 
By stacking the columns of the density matrix $\rho_n$, it can be represented as a vector with $(2LW)^2$ elements; 
this entails to a representation of $\mathcal{F}$ as a matrix of dimension $(2LW)^2 \times (2LW)^2$. 
Due to the matrix dimensions of the Floquet superoperator $\mathcal{F}$ being significantly larger than those of the ordinary Floquet operator $F$, storing the superoperator takes up more memory, limiting the maximum system sizes achievable with this approach. 
However, as we discuss next, for a given system size, the noise-averaged time evolution can be determined directly, leading to a significant reduction in simulation time.

Starting form the initial state $\rho_0 = \ket{\psi_0} \bra{\psi_0}$ and evolving the density matrix as in Eq.~\eqref{eq:Floquet-superoperator-evolution}, we immediately obtain the noise-averaged state $\rho_n = \overline{\ket{\psi_n} \bra{\psi_n}}$. 
This, in turn, gives direct access to the average of the population $\mathcal{P}_n$ defined in Eq.~\eqref{eq:pop-one-seed}:
\begin{equation}
  \label{eq:pop-superop}
  \overline{\mathcal{P}}_n = \overline{|\langle \psi_0| \psi_n \rangle|^2} =
  \bra{\psi_0}\overline{\ket{\psi_n}\bra{\psi_n}}\ket{\psi_0} =
  \braket{\psi_0|\rho_n|\psi_0}.
\end{equation}
We describe how we evaluate the matrix element of the density matrix $\rho_n$ numerically in Appendix~\ref{appx:numerical-superoperator}.

In Fig.~\ref{fig:compare-methods} we compare both methods: that of Eq.~\eqref{eq:pop-superop} and that of numerically averaging Eq.~\eqref{eq:pop-one-seed} over multiple noise realizations. 
The two curves show good agreement, both during the initial exponential decay as well as in the algebraic regime.

\begin{figure}
\includegraphics[width=0.8\columnwidth]{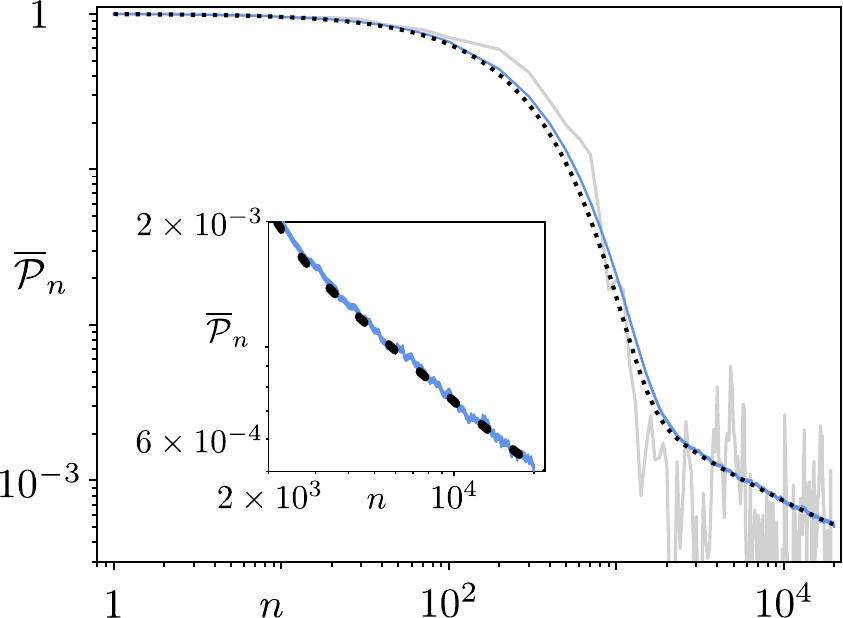}
\caption{
The edge state population computed using the superoperator formalism according to Eq.~\eqref{eq:pop-superop} (dashed black line) agrees well with the one obtained by numerical averaging of Eq.~\eqref{eq:pop-one-seed} over noise realizations (blue). 
The inset zooms in on the algebraic regime. 
For comparison, a single realization of noise is shown in light gray. 
We use $JT/3=1.4$, $JV_t/3=0.05$, $V_o T/3=0.3$, and a system size of $L=24$ and $W=50$ unit cells. 
The blue line is obtained by averaging over 5000 noise realizations, with error bars indicated by line thickness.
}
\label{fig:compare-methods}
\end{figure}

\subsection{All eigenstate populations}
\label{sec:population-brute}

Under the noisy driving, the initially-populated Floquet edge mode mixes with other edge and/or bulk modes, resulting in its population decaying as a function of time. 
To better understand this process, we expand the time-evolved state, $\ket{\psi_n}$, in the orthonormal basis of Floquet eigenstates $\{\ket{\phi_{\alpha}} \}$ defined in Eq.~\eqref{eq:Floquet-states},
\begin{equation}
\ket{\psi_n} = \sum_{\alpha}c_{\alpha, n}\ket{\phi_{\alpha}}.
\label{eq:floquet-basis-expansion}
\end{equation}
Thus, $\mathcal{P}_{\alpha, n} = \lvert c_{\alpha , n} \rvert^2$ are the populations of each of the Floquet eigenstates after $n$ cycles of noisy driving. 
In this notation, when we begin the time evolution with a Floquet edge eigenstate, $\ket{\psi_0}\equiv \ket{\phi_{\alpha = 0}}$, the initial populations are $\mathcal{P}_{\alpha, 0} = \delta_{\alpha, 0}$. 
As the state evolves in time, the weights get redistributed among all the Floquet eigenstates and the populations change. 
After $n$ noisy driving cycles, we obtain the populations from the density matrix $\rho_n$ as in Eq.~\eqref{eq:pop-superop}:
\begin{equation}
  \label{eq:all_averaged_pops}
  \overline{\mathcal{P}}_{\alpha, n} = \braket{\phi_{\alpha} | \rho_n |
    \phi_{\alpha}}.
\end{equation}

\begin{figure*}[tb]
\centering
\includegraphics[width=0.65\textwidth]{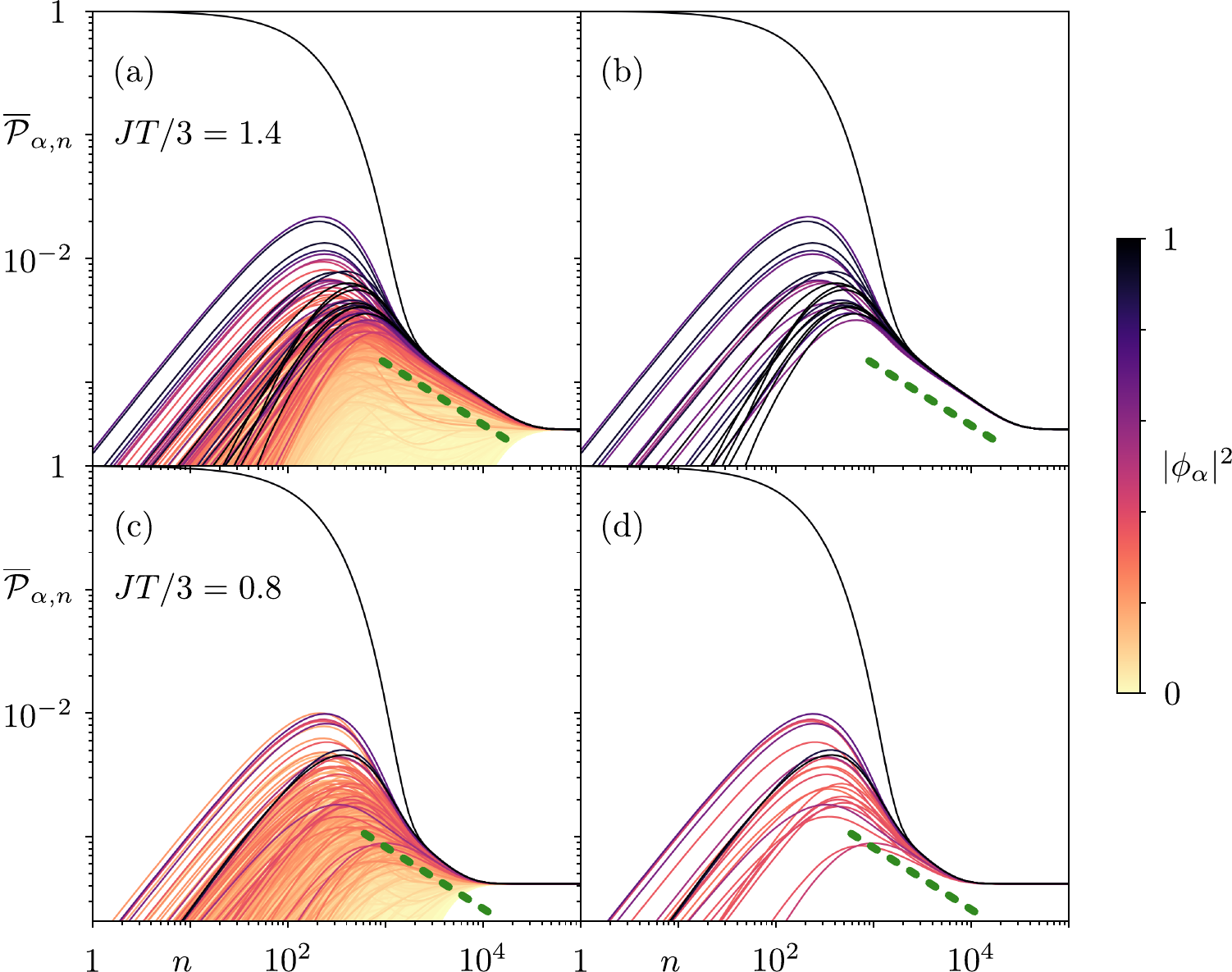}
\caption{
Noise-averaged populations of all Floquet eigenstates [Eq.~\eqref{eq:all_averaged_pops}] in the case of the anomalous Floquet phase [$JT/3=1.4$, panels (a,b)] and in the Chern phase [$JT/3=0.8$, panels (c,d)].
The system size is $L=24$ and $W=50$ unit cells, $JV_t/3=0.05$, and $V_oT/3=0.3$. 
The color scale denotes the probability density summed over the first 3 rows of sites from the bottom edge of the cylinder, such that dark colors indicate Floquet edge states, and light colors indicate bulk states and/or states localized on the top edge. 
The right panels show the populations of those 24 Floquet eigenstates that are closest to the bottom edge. 
The dashed green line shows an algebraic decay $\sim n^{-1/2}$.
}
\label{fig:superop_all_states}
\end{figure*}

In Fig.~\ref{fig:superop_all_states}, we show the populations of all Floquet eigenstates, computed using Eq.~\eqref{eq:all_averaged_pops}, both in the case of the anomalous Floquet phase [$JT/3=1.4$, panels (a,b)], as well as for the Chern phase [$JT/3=0.8$, panels (c,d)].  
The color scale encodes the position of the states, with dark colors denoting modes which are located close to the bottom edge of the cylinder, as is the initial mode $\ket{\psi_0}$. 
As expected, we observe that the noisy driving mixes the initial state with other eigenstates, and that those that are close by in real space become populated first. 
At long times, all populations are equal to the inverse system size: 
The system has converged to an infinite-temperature state. 
However, we observe that the algebraic regime sets in when the edge modes become equally populated, around $n \simeq 2000$ in panels (a,b). 
This suggests that the diffusive behavior is associated to a thermalization only among edge states, which occurs separately from the bulk. 
To better visualize this behavior, we plot the populations of the 24 most edge-localized Floquet eigenmodes in panel (b), a number equal to the circumference of the cylinder. 
It is clear that the algebraic decay coincides with a regime in which the edge mode populations are approximately equal. 
Unfortunately, for the system sizes we are able to reach using the superoperator formalism, this behavior is not so clearly represented for the Floquet-Chern phase, shown in panel (d). 
In the latter case, while it does appear that the edge modes do indeed thermalize separately from the bulk states, there is only a short range of $n$ that is consistent with $\sim n^{-1/2}$ behavior, after which the initial state becomes equally spread among all available Floquet modes.

\begin{figure}
\includegraphics[width=0.8\columnwidth]{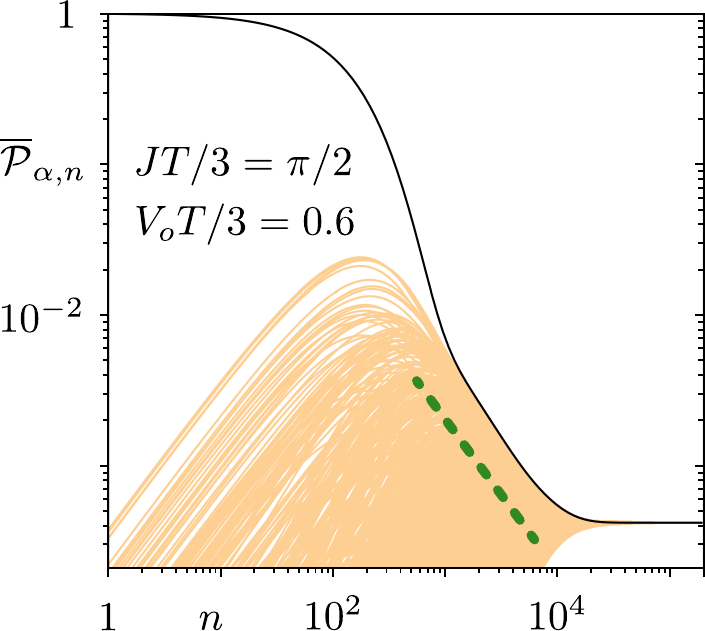}
\caption{
Noise-averaged populations setting $JT/3=\pi/2$ and $V_o T / 3 = 0.6$, for a system consisting on $L=48$ and $W=24$ unit cells. 
The initially-populated state is shown in black, whereas all other states are shown in yellow. 
The dashed green line indicates an algebraic decay $\sim n^{-1}$.
}
\label{fig:bulk_state_decay}
\end{figure}

Based on the above observations, we propose a heuristic explanation for the observed two-stage decay, with exponential decay at short times followed by diffusive decay at later times. 
As we discussed above, in the absence of quenched disorder these two behaviors are dictated by the extended/localized nature of the bulk modes. 
Since states at different momenta do not mix, a localized bulk implies that the initially populated edge state will perform a 1D random walk among states localized nearby, and $\overline{\mathcal{P}}_n \sim n^{-1/2}$. 
This is no longer valid if momentum is not conserved due to quenched disorder: 
Now population can be transferred from the initial edge state to any other edge or bulk state. 
Of course, since noise can affect the state of the system only through local perturbations, transitions to localized bulk states that are far away from the edge or to edge states at the opposite edge are strongly suppressed. 
However, the redistribution of population among edge states that are localized at the same edge is not inhibited. 
Therefore, as generically expected for the decay of a mode into a continuum of modes (formed here by the other modes at the same edge), the population of the initial state decreases exponentially. 
As we observe in Fig.~\ref{fig:superop_all_states}, this exponential decay lasts until all edge modes are equally populated. 
Further decay of the edge state population is due to spreading of the wave function into the bulk. 
For localized bulk states, these dynamics again take the form of a random walk. 
However, since the initial condition of this random walk is a uniformly populated edge, the random walk occurs effectively in one spatial dimension, such that $\overline{\mathcal{P}}_n \sim n^{-1/2}$. 
In contrast, choosing as the initial state a localized state in the bulk, we find 2D diffusive decay with $\overline{\mathcal{P}}_n \sim n^{-1}$. 
This is shown in Fig.~\ref{fig:bulk_state_decay}.

Note that the distinction between 1D and 2D diffusive behavior of edge- and bulk-state decay distinguishes disordered from clean systems. 
In a clean system, momentum conservation implies that for both edge and bulk modes diffusive spreading is restricted to one spatial dimension, leading to $\overline{\mathcal{P}}_n \sim n^{-1/2}$ in both cases. 
Finally, recall that as demonstrated in Fig.~\ref{fig:disordered_cylinder}, while slow diffusive decay results from localized bulk states, the decay of an initially populated edge mode remains exponential if the bulk states are extended.

In our heuristic explanation, we assume that all bulk states are localized or all are extended. 
This assumption does not hold in the Chern phase where, as noted at the end of Sec.~\ref{sec:disordered-case}, not all bulk states can be localized by disorder. 
However, the numerical results shown in Figs.~\ref{fig:disordered_cylinder} and \ref{fig:superop_all_states} demonstrate that diffusive decay of edge states at late times persists also in this case. 
The phenomenological model that we introduce in the following offers an explanation for this surprising behavior.

\section{Phenomenological model}
\label{sec:pheno-model}

To substantiate the above heuristic picture, we develop a phenomenological model for the evolution of populations in disordered systems. 
This model is based on a semiclassical Floquet master equation for the populations of Floquet states, which we derive from the exact dynamics. 
In the semiclassical Floquet master equation, the transfer of population from a Floquet state $\alpha$ to a state $\beta$ during one driving cycle is described by the element $\mathcal{W}_{\alpha \to \beta}$ of the population transfer matrix. 
Phenomenological assumptions about the form of these matrix elements enable an analytical understanding of the dynamics.

\subsection{Semiclassical Floquet master equation}
\label{sec:semicl-floq-mast}

The time evolution of the density matrix $\rho_n$ is essentially due to two distinct processes: 
(i) Weak noise induces transitions between Floquet states and thus leads to a slow redistribution of state populations given by the diagonal elements $\mathcal{P}_{\alpha, n} = \braket{\phi_{\alpha} | \rho_n | \phi_{\alpha}}$. 
(In this section, we do not indicate the average over noise explicitly; all populations are understood to be averaged.) 
(ii) Coherent dynamics cause off-diagonal elements $\braket{\phi_{\alpha} | \rho_n | \phi_{\beta}}$ with $\alpha \neq \beta$ to dephase as $\sim e^{i \left( \varepsilon_{\alpha} - \varepsilon_{\beta} \right) n T}$.
Assuming that dephasing of coherences is much faster than the redistribution of populations---as we demonstrate below, this assumption is met for sufficiently strong disorder---we describe the density matrix $\rho_n$ approximately as an incoherent mixture of Floquet states,
\begin{equation}
  \label{eq:rho-n-diagonal}
  \rho_n = \sum_{\alpha} \mathcal{P}_{\alpha,n} \ket{\phi_{\alpha}} \bra{\phi_{\alpha}},
\end{equation}
and we project the evolution Eq.~\eqref{eq:Floquet-superoperator-evolution} onto a semiclassical Floquet master equation for the populations $\mathcal{P}_{\alpha,n}$,
\begin{equation}
  \label{eq:P-G-P}
  \mathcal{P}_{\alpha, n + 1} = \sum_{\beta} \mathcal{G}_{\alpha, \beta} \mathcal{P}_{\beta, n},
\end{equation}
where,
\begin{equation}\label{eq:Gmat-definition}
  \mathcal{G}_{\alpha, \beta} = \braket{\phi_{\alpha} |
    \mathcal{F}(\ket{\phi_{\beta}} \bra{\phi_{\beta}}) | \phi_{\alpha}}.
\end{equation}
In the absence of timing noise, the time-evolution superoperator reduces to $\mathcal{F} \rho = F \rho F^{\dagger}$, with $F$ given by Eq.~\eqref{eq:Floquet}. For the Floquet states, it follows from
Eq.~\eqref{eq:Floquet-states} that
\begin{equation}
  \mathcal{G}_{\alpha, \beta} = \lvert \braket{\phi_{\alpha} | \phi_{\beta}} \rvert^2 =
  \delta_{\alpha, \beta},
\end{equation}
meaning that there is no redistribution of populations in the absence of noise. 
We define the matrix $\mathcal{W}_{\beta \to \alpha}$ that describes the noise-induced transfer of population from the state $\beta$ to $\alpha$ as the deviation from the noiseless limit:
\begin{equation}
  \label{eq:population-transfer-matrix}
  \mathcal{W}_{\beta \to \alpha} = \mathcal{G}_{\alpha, \beta} - \delta_{\alpha, \beta}.
\end{equation}
As we show in Appendix~\ref{appx:semicl-floq-mast-app}, a sufficient condition for the semiclassical Floquet master equation~\eqref{eq:P-G-P} to describe noise-induced heating to infinite temperature is that the population transfer matrix is symmetric, $\mathcal{W}_{\alpha \to \beta} = \mathcal{W}_{\beta \to \alpha}$. 
Then, the master equation can be recast as
\begin{equation}
  \label{eq:semiclassical-Floquet-master-equation-alternative}
  \mathcal{P}_{\alpha, n + 1} = \mathcal{P}_{\alpha, n} + \sum_{\beta \neq \alpha} \mathcal{W}_{\beta
    \to \alpha} \left(  \mathcal{P}_{\beta, n} - \mathcal{P}_{\alpha, n} \right).
\end{equation}
This equation describes how the population of state $\alpha$ increases during a noisy driving cycle due to transitions $\beta \to \alpha$, which occur with probability $\mathcal{W}_{\beta \to \alpha} \mathcal{P}_{\beta, n}$, and decreases due to transitions $\alpha \to \beta$, which occur with probability $\mathcal{W}_{\alpha \to \beta} \mathcal{P}_{\alpha, n} = \mathcal{W}_{\beta \to \alpha} \mathcal{P}_{\alpha, n}$.

\subsection{Comparison with numerical results}
\label{sec:Wmat-from-numerics}

We test the above assumptions by numerically constructing the population transfer matrix using Eqs.~\eqref{eq:Gmat-definition} and \eqref{eq:population-transfer-matrix} [the matrix elements in Eq.~\eqref{eq:Gmat-definition} can be obtained from the vectorized form of the Floquet superoperator as detailed in Appendix~\ref{appx:numerical-superoperator}] for a system in the anomalous phase [same parameters as in Fig.~\ref{fig:superop_all_states}(a, b)]. 
The conservation of probability, $\sum_{\alpha} \mathcal{W}_{\beta \to \alpha} = 0$ [see Eq.~\eqref{eq:conservation-of-probability}], is preserved up to deviations in the range of $10^{-16}$ to $10^{-11}$, which sets the scale of our numerical errors. 
Furthermore, also the assumption that $\mathcal{W}$ is symmetric is borne out by the numerics, with $|| \mathcal{W} - \mathcal{W}^T || / ||\mathcal{W}|| \simeq 3\cdot 10^{-5} $, where $|| \cdot ||$ is the Euclidean norm.

\begin{figure*}[tb]
\centering
\includegraphics[width=0.65\textwidth]{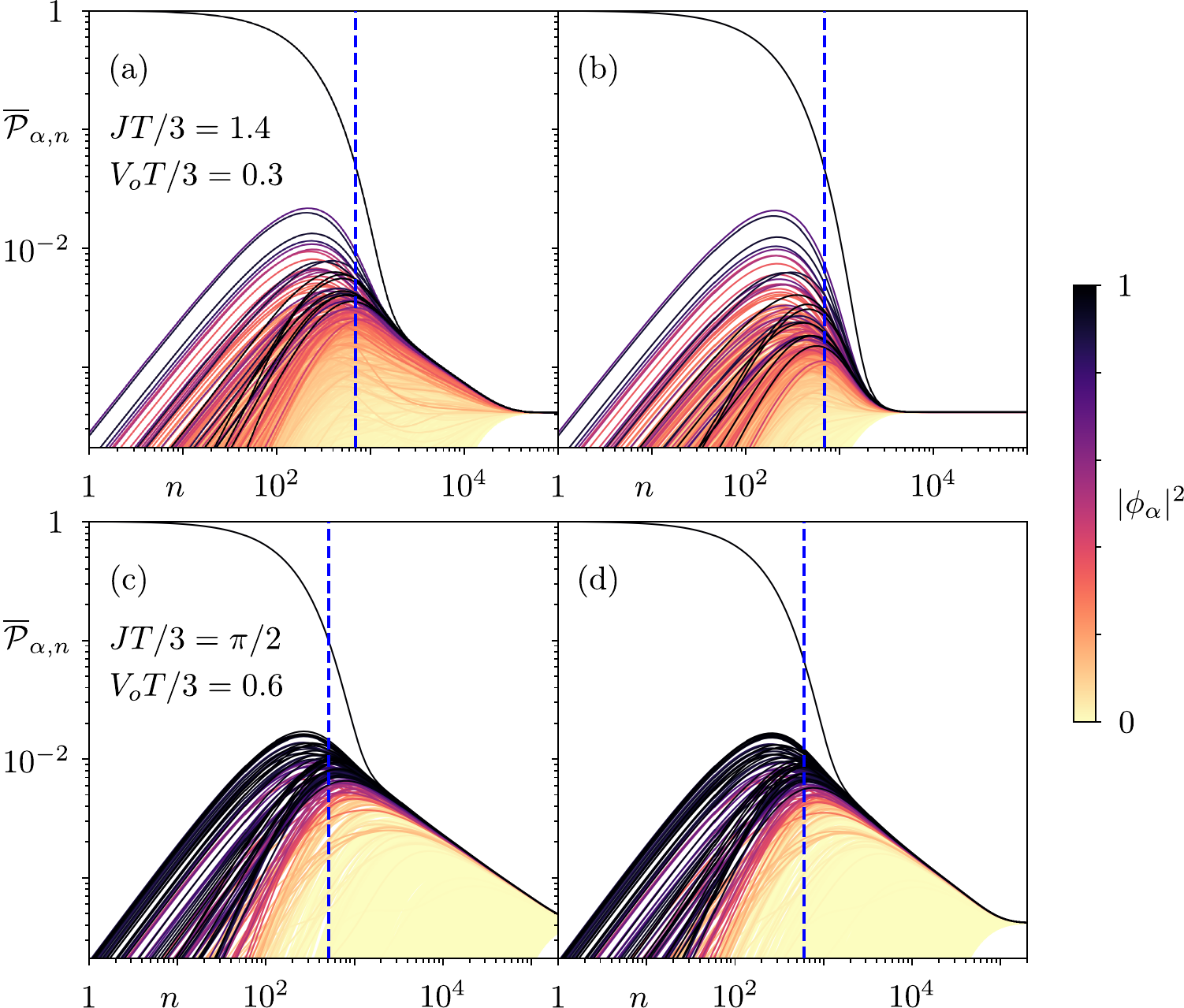}
\caption{
Noise-averaged populations at $JT/3=1.4$, $V_oT/3=0.3$ [panels (a,b)]   and at $JT/3=\pi/2$, $V_oT/3=0.6$ [panels (c,d)]. 
In the former case, there is a poor agreement between the superoperator calculation (left) and the semiclassical Floquet master equation (right). 
For the higher disorder strength, the agreement is improved, though the semiclassical Floquet master equation reaches the infinite-temperature state faster than the superoperator.
The system size, noise strength, and the color scale are the same as in Fig.~\ref{fig:superop_all_states}. 
The vertical dashed lines indicate our estimate for the edge thermalization time Eq.~\eqref{eq:edge-thermalization-time}.
}
\label{fig:superop_vs_Wmat}
\end{figure*}

Having obtained the population transfer matrix, we iterate Eq.~\eqref{eq:semiclassical-Floquet-master-equation-alternative} to construct the Floquet eigenstate populations. 
In the anomalous phase, $JT/3=1.4$ and $V_oT/3=0.3$, the semiclassical Floquet master equation shows a poor agreement with the superoperator calculation, as shown in Fig.~\ref{fig:superop_vs_Wmat}(a, b). 
In particular, the semiclassical Floquet master equation fails to capture the algebraic decay of the populations at intermediate times. 
When we increase the disorder strength, setting $JT/3=\pi/2$ and $V_oT/3=0.6$, we observe a much better agreement, see Fig.~\ref{fig:superop_vs_Wmat}(c, d). 
We attribute this change to the fact that the rate at which off-diagonal elements are generated in the density matrix must contain a matrix element of one of the Hamiltonians (that is, a local operator) between different Floquet states, and these matrix elements will be smaller when the Floquet states are more strongly localized. 
Also, the dephasing of off-diagonal elements of the density matrix occurs even without noise simply due to quasi-energy differences, and these quasi-energy differences are larger for stronger disorder (at least for states that are close by in real space). 
So the approximation we make in setting off-diagonal elements to zero in the semiclassical Floquet master equation should be better for stronger disorder.

In addition, since the approximation consists in making the density matrix more diagonal (less off-diagonal) by hand, and in this sense brings it closer to the infinite-temperature steady state, it appears reasonable that the diffusive regime is shortened. 
This is indeed visible in Fig.~\ref{fig:superop_vs_Wmat}, where in both parameter regimes the infinite-temperature state is reached more quickly in the semiclassical Floquet master equation (panels b, d) than in the corresponding superoperator calculations (a, c).

\subsection{Phenomenological population transfer matrix}
\label{sec:population-transfer-matrix}

Having established the validity of the diagonal approximation in Eq.~\eqref{eq:rho-n-diagonal} for sufficiently strong disorder, our next aim is to obtain deeper analytical insights into the dynamics described by the semiclassical Floquet master equation~\eqref{eq:semiclassical-Floquet-master-equation-alternative}. 
To that end we propose a simplified phenomenological description of the system. 
Instead of a continuous distribution of localization lengths of bulk states as we observe in the full numerics for disordered systems, we consider only two types of bulk states: localized and extended. 
We denote these types of states by $\ket{\phi_b}$ and $\ket{\phi_d}$, respectively. 
Edge states are denoted by $\ket{\phi_e}$. 
We assume these states to have the following general structure:
\begin{equation}
  \label{eq:edge-bulk-state-scaling}
  \begin{split}  
    \braket{\mathbf{r} | \phi_e} & \sim \frac{1}{\sqrt{L}} e^{- y/\xi}, \\
    \braket{\mathbf{r} | \phi_b} & \sim e^{ - \vert \mathbf{r} - \mathbf{r}_b \vert / \xi}, \\
    \braket{\mathbf{r} | \phi_d} & \sim \frac{1}{\sqrt{LW}},
  \end{split}
\end{equation}
with $\mathbf{r} = \left( x, y \right)$ the position vector characterizing a lattice site, $\xi$ the localization length, and $L$, $W$ the linear dimensions of the system. 
Each localized bulk state is assumed to be centered at a position $\mathbf{r}_b$. 
For simplicity, we consider a square lattice geometry with one lattice site per unit cell. 
This simplifying assumption does not affect our results qualitatively. 
Furthermore, we assume the localization length to be small: $\xi \ll 1$, with distances measured in units of the lattice spacing.

The forms of the states in Eq.~\eqref{eq:edge-bulk-state-scaling} directly imply how the elements of the population transfer matrix depend on the parameters of the model. 
To quadratic order in $V_t$, $\mathcal{W}_{\alpha \to \beta}$ is proportional to the absolute value squared of the matrix element of a local operator $\sim H_m$ between the states $\ket{\phi_{\alpha}}$ and $\ket{\phi_{\beta}}$,
\begin{equation}
  \mathcal{W}_{\alpha \to \beta} \sim \left( V_t \sum_{\mathbf{r}}
    \braket{\mathbf{r} | \phi_{\alpha}} \braket{\mathbf{r} | H_m |
      \phi_{\beta}} \right)^2.
\end{equation}
The appearance of $H_m$ implies that the overall scaling with hopping amplitude and noise strength is $\mathcal{W}_{\alpha \to \beta} \sim \left( J V_t \right)^2$. 
To understand how $\mathcal{W}_{\alpha \to \beta}$ depends on system size and the distance between localized bulk states, we note that the quantities $\braket{\mathbf{r} | \phi_{\alpha}} \braket{\mathbf{r} | H_m | \phi_{\beta}}$ can, to a first approximation, be regarded as having random phases that are uncorrelated between different lattice sites $\mathbf{r}$. 
As a result, the sum scales as $\sqrt{N}$, where $N$ is the number of terms in the sum.

With this in mind, we first consider the population transfer matrix element between two edge states $e$ and $e^{\prime}$. 
For a sufficiently short localization length $\xi$, the sum over lattice sites $\mathbf{r}$ is reduced to a sum along the edge and thus runs over $L$ terms. 
Therefore, accounting for the factor $L^{-1/2}$ in Eq.~\eqref{eq:edge-bulk-state-scaling}, we obtain
\begin{equation}
  \label{eq:Wee}
  \mathcal{W}_{e \to e^{\prime}} \sim \left( J V_t \right)^2 / L. 
\end{equation}
Note that this form captures only the overall scaling of the matrix element with system parameters; 
the precise value of the matrix element varies depending on the choice of states $e$ and $e'$. For transitions between an edge and a localized bulk state, the sum over $\mathbf{r}$ collapses to a single term with $\mathbf{r} = \mathbf{r}_b$, and we find
\begin{equation}
  \label{eq:Web}
  \mathcal{W}_{e \to b} \sim \left( J V_t \right)^2 e^{- 2 y_b/\xi}/L.
\end{equation}
Finally, for a pair of localized bulk states, the sum collapses to two terms that are exponentially small in the distance between the states,
\begin{equation}
  \label{eq:Wbb}
  \mathcal{W}_{b \to b^{\prime}} \sim \left( J V_t \right)^2 e^{ - 2
    \vert \mathbf{r}_b - \mathbf{r}_{b^{\prime}}\vert / \xi }.
\end{equation}
Similar arguments lead to the matrix elements for transitions involving extended bulk states:
\begin{equation}
  \label{eq:Wed}
  \mathcal{W}_{e \to d} \sim  \left( J V_t \right)^2 e^{-2/\xi}/\left( LW \right),
\end{equation}
and
\begin{equation}
  \label{eq:Wbd-Wdd}
  \mathcal{W}_{b \to d}, \, \mathcal{W}_{d \to d'} \sim  \left( J V_t \right)^2
  / \left( LW \right).
\end{equation}
In the following, we discuss the time evolution described by the semiclassical Floquet master equation~\eqref{eq:semiclassical-Floquet-master-equation-alternative} with the phenomenological forms of the populations transfer matrix elements given above. 
We focus on the different regimes of the decay of an edge state, the initial exponential and the subsequent diffusive decay, and we address the question, under which conditions diffusion persists in the presence of extended bulk states.

\subsection{Short-time dynamics: edge thermalization}
\label{sec:edge-thermalization}

The phenomenological population transfer matrix elements in Eqs.~\eqref{eq:Wee}, \eqref{eq:Web}, and~\eqref{eq:Wed} indicate that at short times, the decay of the initially populated edge state is mainly due to transitions to other edge states, whereas transitions to bulk states are suppressed by a factor of order $O \! \left( e^{-2/\xi} \right)$. 
To describe the resulting thermalization among edge modes approximately, we set the populations of bulk states to zero, $\mathcal{P}_{b, n} = \mathcal{P}_{d, n} = 0$, and we assume that the populations of all edge modes at the same edge as the initial edge mode $e_1$ are the same, $\mathcal{P}_{e_2, n} = \mathcal{P}_{e_3, n} = \dotsb = \mathcal{P}_{e_L, n}$.
Then, the semiclassical Floquet master equation~\eqref{eq:semiclassical-Floquet-master-equation-alternative} can be rewritten in terms of the difference of populations, $\Delta_n = \mathcal{P}_{e_{1}, n} - \mathcal{P}_{e_{2}, n}$, as 
\begin{equation}  
  \Delta_{n + 1} = \left( 1 - w_e \right) \Delta_n,
\end{equation}
where $w_e$ is the average of $L \mathcal{W}_{e_1 \to e'}$ over $e' \neq e_1$ and we dropped terms of order $O(1/L)$. 
The solution to this recursion relation with initial condition $\Delta_0 = 1$ reads
\begin{equation}
  \label{eq:Delta-n}
  \Delta_n = \left( 1 - w_e \right)^{n} = e^{- n/\tau_e},
\end{equation}
with the time scale on which the difference between the populations of edge states decays given by
\begin{equation}
  \label{eq:edge-thermalization-time}
  \tau_e = -1/\ln(1 - w_e) \sim 1/
  w_e.
\end{equation}
Note that according to Eq.~\eqref{eq:Wee}, $w_e$, defined as the average of $L \mathcal{W}_{e_1 \to e'}$, and thus also the edge thermalization time $\tau_e$ do not depend on the number of edge states $L$. 
The populations of individual edge states can be obtained from Eq.~\eqref{eq:Delta-n} by using $\mathcal{P}_{e_1, n} + \left( L - 1 \right) \mathcal{P}_{e_2, n} = 1$.

Our estimate of the edge thermalization time Eq.~\eqref{eq:edge-thermalization-time} is in good agreement with our numerical simulations. 
In Fig.~\ref{fig:superop_vs_Wmat}, we indicate $\tau_e \sim 1/w_e$ by a vertical dashed line. 
To find $w_e$, we use that the evolution during the first driving cycle, described by Eq.~\eqref{eq:semiclassical-Floquet-master-equation-alternative} with $n = 0$, yields $\mathcal{P}_{e', 1} = \mathcal{W}_{e_1 \to e'}$ for $e' \neq e_1$.
Therefore, $w_e$ is obtained by averaging $L \mathcal{P}_{e', 1}$ over $e'$. 
Specifically, since $L=24$ in our simulations, we average the populations of the 23 states closest to the edge, excluding the initial state.

\subsection{Diffusive decay at long times}

During edge thermalization, the population of states remains localized at the boundary of the system, but extends homogeneously along the edge. 
The subsequent spreading of population into the bulk can thus be regarded as occurring only along the direction orthogonal to the edge, while the population remains homogeneous in the direction parallel to the edge. 
This results in 1D diffusive motion (as we show below, diffusion persists also in the presence of a sufficiently small number of extended bulk states). 
For 1D diffusion, the decay of an initially localized population follows the same power-law dependence, $\sim n^{-1/2}$, regardless of whether the population is initially localized at the boundary or in the bulk of the system~\cite{Rieder2018, Sieberer2018}. 
Therefore, to describe these dynamics approximately, we make the following simplifying assumptions: 
(i)~We consider a system with periodic boundary conditions, consisting of $\left( 1 - \nu_d \right) L W$ localized bulk states, and $\nu_d L W$ extended bulk states; 
that is, the total number of states is $L W$, and $\nu_d$ is the fraction of extended bulk states. 
(ii)~The populations of bulk states are homogeneous in one spatial direction, meaning that the population of a bulk state localized at $\mathbf{r}_b = \left( x, y \right)$ depends only on $y$, $\mathcal{P}_{b, n} = \mathcal{P}_{y, n}$. 
(iii)~The populations of extended bulk states, $\mathcal{P}_{d, n}$, are all equal. 
Initially, extended bulk states are not populated, $\mathcal{P}_{d, 0} = 0$.

Under these assumptions, the total population of localized and extended bulk states is given by
\begin{equation}
  \sum_b \mathcal{P}_{b, n} + \sum_d \mathcal{P}_{d, n} = L \sum_y
  \mathcal{P}_{y, n} + \nu_d L W \mathcal{P}_{d, n} = 1.
\end{equation}
With this relation, the semiclassical Floquet master equation~\eqref{eq:semiclassical-Floquet-master-equation-alternative} for extended bulk states can be written as
\begin{equation}
  \label{eq:P-d}
  \mathcal{P}_{d, n + 1} = \mathcal{P}_{d, n} + w_{b, d} \left( \frac{1}{L W} - \mathcal{P}_{d, n}
  \right).
\end{equation}
Here we further assume that averaging $L W \mathcal{W}_{b \to d}$ over localized or extended bulk states yields the same result, which we denote by $w_{b, d}$.
The solution of Eq.~\eqref{eq:P-d} that obeys the initial condition $\mathcal{P}_{d, 0} = 0$ reads
\begin{equation}
  \mathcal{P}_{d, n} = \frac{1}{L W} \left[ 1 - \left( 1 - w_{b, d} \right)^n
  \right].
\end{equation}

Next, we consider the evolution of localized bulk states. 
If the localization length is short, $\xi \ll 1$, only transitions to neighboring lattice sites have to be taken into account, for which we set $\mathcal{W}_{b \to b'} = w_b e^{-2/\xi}$. 
We thus obtain
\begin{multline}
  \label{eq:P-y-nearest-neighbor}
  \mathcal{P}_{y, n + 1} = \mathcal{P}_{y, n} + w_b e^{-2/\xi} \left( \mathcal{P}_{y + 1,
      n} + \mathcal{P}_{y - 1, n} - 2 \mathcal{P}_{y, n} \right) \\ + w_{b, d}
  \nu_d \left( \mathcal{P}_{d, n} - \mathcal{P}_{y, n} \right).
\end{multline}
This equation can be solved by taking the discrete Fourier transform of $\mathcal{P}_{y, n}$ with respect to $y$. 
Then, for $n \to \infty$ and weak noise, $w_{b, d} \ll 1$, we find
\begin{equation}
  \label{eq:diffusive-exponential}
  \mathcal{P}_{y = 0, n} \sim \frac{1}{2 L \sqrt{\pi w_b n}} e^{- w_{b, d}
    \nu_d n + 1/\xi}.
\end{equation}
That is, the population that is initially localized at $y = 0$ decays first algebraically, $\mathcal{P}_{y = 0, n} \sim n^{-1/2}$. 
However, at $n_{\mathrm{exp}} \sim 1/(w_{b, d} \nu_d)$, this slow diffusive decay is cut off by the exponential dependence $\mathcal{P}_{y = 0, n} \sim \e^{- w_{b, d} \nu_d n}$.

In the anomalous Floquet-Anderson phase, the bulk of the system is fully localized by disorder. 
To describe this phase, we set $\nu_d = 0$ in the phenomenological model. 
Then, Eq.~\eqref{eq:diffusive-exponential} describes purely diffusive decay in qualitative agreement with our numerical simulations. 
In contrast, there is always a finite number of extended modes in the Floquet-Chern phase, meaning that $\nu_d \neq 0$. 
But also in this case, we numerically observe diffusive and not exponential decay at late times. 
How can this observation be reconciled with Eq.~\eqref{eq:diffusive-exponential}?

The exponential decay implied by Eq.~\eqref{eq:diffusive-exponential} is not visible if the stationary state with equal populations of all Floquet modes, $\mathcal{P}_{y, n} = \mathcal{P}_{d, n} = 1/(L W)$, is reached before $n_{\mathrm{exp}}$. 
Assuming the decay to be purely diffusive, the steady state is reached at the Thouless time, $n_{\mathrm{Th}} = W^2 e^{2/\xi}/(4 \pi w_b)$.
If there is a finite fraction of extended bulk modes in the thermodynamic limit, then $n_{\mathrm{exp}} \ll n_{\mathrm{Th}}$, and exponential decay will always be visible. 
However, if there is only a finite \emph{number} of extended bulk states such that $\nu_d \sim 1/(L W)$, then a reversal of the order of time scales is possible, $n_{\mathrm{Th}} \lesssim n_{\mathrm{exp}}$. 
This will be the case, in particular, when $\xi$ is not too small such that $e^{2/\xi}$ is of order one. 
Note that in deriving Eq.~\eqref{eq:P-y-nearest-neighbor}, we only took into account transitions to neighboring localized states. 
This is no longer justified if $\xi$ is not small. 
However, including transitions to more distant states will not change the qualitative behavior described by Eq.~\eqref{eq:diffusive-exponential}, but will only lead to a modification of parameters that further reduces $n_{\mathrm{Th}}$.

We thus conclude that the Chern phase is better described by setting $\nu_d \sim 1/(L W)$ in our phenomenological model. 
However, we reiterate that our assumption of there being only two types of states (localized and extended) is a significant simplification. 
In the full microscopic model, there is a continuous distribution of localization lengths.

\section{Conclusions}
\label{sec:conclusions}

In this work, we presented a detailed study of various Floquet topological phases in the presence of both quenched disorder as well as timing noise.  
We considered the two most common 2D Floquet topological phases, the anomalous Floquet-Anderson phase, and Floquet-Chern phase, and presented an in-depth analysis of the edge state decoherence into the bulk.

Decoherence is interesting from an experimental standpoint because of the impossibility of obtaining perfectly-periodic driving. 
In the case of photonic crystal experiments \cite{Rechtsman2013, Maczewsky2017, Mukherjee2017}, timing noise could be due to non-periodic variations in the distances between the waveguides along the light propagation direction. 
For ultracold atoms in optical lattices \cite{Wintersperger2020}, decoherence is due to the inherent imperfections in laser light sources. 
As a result, the chiral edges will inevitably leak into the bulk.

To understand the effects of noise, we studied the population of an initially filled edge state as a function of time. 
In a clean system, the behavior is similar to a 1D case due to translational invariance. 
Consequently, the anomalous Floquet topological phase at resonant driving conditions shows a power law decay, as does the Chern phase provided that we choose a momentum at which bulk states are non-dispersing (see Fig.~\ref{fig:clean_ribbon}). 
In the presence of dispersing, delocalized bulk states, however, the edge mode shows an exponential decay.

When quenched disorder is included, momentum is no longer a good quantum number, and the edge mode population shows surprising features. 
We found that it decays exponentially at short times, both in the case of a Floquet-Chern phase as well as in the Floquet-Anderson phase. 
However, both phases show an additional diffusive regime after the initial exponential decay (see Fig.~\ref{fig:disordered_cylinder}). 
To understand this better, we looked at the populations of all the Floquet eigenstates, and found that the diffusive regime occurs once all the edge states have thermalized amongst themselves (Figs.~\ref{fig:superop_all_states} and \ref{fig:superop_vs_Wmat}). 
Remarkably, this thermalization time remains finite even as the number of edge modes diverges in the thermodynamic limit.

In an experimental setup where it is desired to have a long-lasting edge mode, such as signal processing using the edge states or information transmission through edge states, the edge states have to survive even with timing noise.
Our results show that to build such a platform, it is suitable to tune them into an anomalous Floquet topological phase at resonant driving and to minimize quenched disorder. 
However, resonant driving may not always be possible to achieve. 
In those cases, one can rely on the presence of quenched disorder, even purposely introducing disorder if necessary, in order to reduce the decay rate of the edge mode.

\begin{acknowledgments}
The authors would like to thank Ulrike Nitzsche for technical support. 
This   work was supported by the Deutsche Forschungs Gemeinschaft (DFG, German Research Foundation) under Germany’s Excellence Strategy through the W\"urzburg-Dresden Cluster of Excellence on Complexity and Topology in Quantum Matter–ct.qmat(EXC 2147, project-id 390858490), as well as through the DFG Grant FU 1253/1-1. 
L.S.\ acknowledges support from the Austrian Science Fund (FWF) through the projects 10.55776/P33741 and 10.55776/COE1, and from the European Union - NextGenerationEU.
\end{acknowledgments}

\appendix

\section{Floquet superoperator}
\label{appx:numerical-superoperator}

To find a matrix representation of the Floquet superoperator that can be employed in numerical simulations, we rewrite Eq.~\eqref{eq:Floquet-superoperator-evolution} in vectorized form. 
This is done by stacking the columns of $\rho_n$ to form a vector $\kket{\rho_n}$. 
Under this operation, the product of three matrices $A$, $B$, and $C$ becomes
\begin{equation}
  \label{eq:vectorization}
  \kket{A B C} = C^{\transpose} \otimes A \kket{B},
\end{equation}
where $\otimes$ is the Kronecker product of matrices.

According to Eq.~\eqref{eq:Floquet-superoperator}, the Floquet superoperator is the product of functions of the superoperators $\mathcal{H}_m$ defined in Eq.~\eqref{eq:H-m-superoperator}, $\mathcal{F} = f(\mathcal{H}_3) f(\mathcal{H}_2) f(\mathcal{H}_1)$, where
\begin{equation}
  f(\varepsilon) = e^{-i \varepsilon T/3} \sinc(\varepsilon V_t/3).
\end{equation}
A matrix representation of each of the factors $f(\mathcal{H}_m)$ can be obtained by diagonalizing the superoperators $\mathcal{H}_m$. 
The vectorized form of the latter can be obtained straightforwardly using Eq.~\eqref{eq:vectorization}:
\begin{equation}
  \kket{\mathcal{H}_m \rho} = \kket{H_m \rho \id - \id \rho H_m} = \left( \id
    \otimes H_m - H_m^{\transpose} \otimes \id \right) \kket{\rho}.
\end{equation}
We denote by $O_m$ the matrix that diagonalizes $H_m$,
\begin{equation}
  O_m^{\dagger} H_m O_m = D_m,
\end{equation}
where $D_m$ is a diagonal matrix. 
Since the Hamiltonians $H_m$ describe hopping across disconnected bonds, they consist of $2 \times 2$ blocks that can be diagonalized by hand, meaning that $O_m$ and $D_m$ can be found analytically. 
Then, with $\mathcal{O}_m = O_m^{*} \otimes O_m$,
\begin{multline}  
     \mathcal{O}_m^{\dagger} \mathcal{H}_m \mathcal{O}_m \\
     \begin{aligned}
       = & \left( O_m^{\transpose}
      \otimes O_m^{\dagger} \right) \left( \id \otimes H_m - H_m^{\transpose}
      \otimes \id \right) \left( O_m^{*} \otimes O_m \right) \\ 
    = & \id\otimes D_m - D_m \otimes \id = \mathcal{D}_m,
     \end{aligned}
\end{multline}
where $\mathcal{D}_m$ is again a diagonal matrix. 
Each of the factors in $\mathcal{F}$ can thus be written as
\begin{equation}
  f(\mathcal{H}_m) = \mathcal{O}_m f(\mathcal{D}_m) \mathcal{O}_m^{\dagger}.
\end{equation}

Vectorization also offers a convenient way to evaluate the noise-averaged population of the edge state in Eq.~\eqref{eq:pop-superop}. 
Denoting by $\kket{\psi_0}$ the vectorized form of the projector $\ket{\psi_0} \bra{\psi_0}$, we find
\begin{equation}
  \overline{\mathcal{P}}_n = \braket{\psi_0 | \rho_n | \psi_0} =
  \tr(\ket{\psi_0} \bra{\psi_0} \rho_n) = \bbraket{\psi_0|\rho_n}.
\end{equation}

\section{Semiclassical Floquet master equation}
\label{appx:semicl-floq-mast-app}

To derive Eq.~\eqref{eq:semiclassical-Floquet-master-equation-alternative}, we start from Eq.~\eqref{eq:P-G-P}, where we express $\mathcal{G}_{\alpha, \beta}$ in terms of the population transfer matrix introduced in Eq.~\eqref{eq:population-transfer-matrix}, which leads to
\begin{equation}
  \label{eq:semiclassical-Floquet-master-equation}
  \mathcal{P}_{\alpha, n + 1} = \mathcal{P}_{\alpha, n} + \sum_{\beta} \mathcal{W}_{\beta \to \alpha}
  \mathcal{P}_{\beta, n}.
\end{equation}
This evolution equation conserves the normalization of populations,
$\sum_{\alpha} \mathcal{P}_{\alpha, n} = 1$, if
\begin{equation}
  \sum_{\alpha, \beta} \mathcal{W}_{\beta \to \alpha} \mathcal{P}_{\beta, n} = 0,
\end{equation}
which has to hold for all $\mathcal{P}_{\beta, n} > 0$. Therefore,
\begin{equation}
  \sum_{\alpha} \mathcal{W}_{\beta \to \alpha} = 0,
  \label{eq:conservation-of-probability}
\end{equation}
which in turn implies
\begin{equation}
  \label{eq:W-alpha-to-alpha}
  \mathcal{W}_{\alpha \to \alpha} = - \sum_{\beta \neq \alpha} \mathcal{W}_{\alpha \to \beta}.
\end{equation}
This condition allows us to recast the semiclassical Floquet master equation in the following form:
\begin{equation}
\label{eq:evolution-pheno}
  \mathcal{P}_{\alpha, n + 1} = \mathcal{P}_{\alpha, n} + \sum_{\beta \neq \alpha} \left( \mathcal{W}_{\beta
      \to \alpha} \mathcal{P}_{\beta, n} - \mathcal{W}_{\alpha \to \beta} \mathcal{P}_{\alpha, n} \right).
\end{equation}
Classical noise is generically expected to cause heating. 
Therefore, we expect the state at infinite temperature with $\mathcal{P}_{\alpha, \infty} = 1/D$, where $D = 2 L W$ is the Hilbert space dimension, to be the steady state.
Inserting $\mathcal{P}_{\alpha, \infty}$ in Eq.~\eqref{eq:semiclassical-Floquet-master-equation}, we find that $\mathcal{P}_{\alpha, \infty}$ is a steady state if
\begin{equation}
  \sum_{\beta} \mathcal{W}_{\beta \to \alpha} = 0.
\end{equation}
Due to Eq.~\eqref{eq:conservation-of-probability}, a sufficient condition for this relation to hold is that $\mathcal{W}_{\alpha \to \beta} = \mathcal{W}_{\beta \to \alpha}$ is symmetric. 
Under the assumption that this is indeed the case, Eq.~\eqref{eq:semiclassical-Floquet-master-equation-alternative} follows immediately from Eq.~\eqref{eq:evolution-pheno}.

\bibliography{references}

\end{document}